# Ferroelectricity-tuned band topology and superconductivity in two-dimensional materials and related heterostructures

Jianyong Chen, Ping Cui, Zhenyu Zhang

Ferroelectricity, band topology, and superconductivity are respectively local, global, and macroscopic properties of quantum materials, and understanding their mutual couplings offers unique opportunities for exploring rich physics and enhanced functionalities. In this mini-review, we attempt to highlight some of the latest advances in this vibrant area, focusing in particular on ferroelectricity-tuned superconductivity and band topology in two-dimensional (2D) materials and related heterostructures. We will first present results from predictive studies of the delicate couplings between ferroelectricity and topology or superconductivity based on first-principles calculations and phenomenological modeling, with ferroelectricity-enabled topological superconductivity as an appealing objective. Next, we will cover the latest advances on experimental studies of ferroelectricity-tuned superconductivity based on different 2D materials or van der Waals heterostructures. Finally, as perspectives, we will outline schemes that may allow to materialize new types of 2D systems that simultaneously harbor ferroelectricity and superconductivity, or that may lead to enhanced ferroelectric superconductivity, ferroelectric topological superconductivity, and new types of superconducting devices such as superconducting diodes.

J. Chen

Institute of Quantum Materials and Physics, Henan Academy of Sciences

Zhengzhou, 450046, China

Email: chenjy@hnas.ac.cn

P. Cui, Z. Zhang

International Center for Quantum Design of Functional Materials (ICQD), Hefei National Research Center for Physical Sciences at the Microscale, University of Science and Technology of China

Hefei, 230026, China

Emails: cuipg@ustc.edu.cn; zhangzy@ustc.edu.cn

# 1. Introduction

Ferroelectricity, band topology, and superconductivity are fundamental concepts in condensed matter physics, each playing a crucial role in understanding the properties of materials. Ferroelectricity refers to the ability of certain materials exhibiting non-volatile switchable electric polarizations induced by spontaneous inversion-symmetry breaking. Band topology is a global property of occupied bands in materials and is closely associated with electronic orbital interaction. Superconducting pairing is intricately related to the geometry of the Fermi surface in metallic systems and electron-electron correlations. The modern theory of polarization is rooted in a direct connection between electric polarization and topological quantum phase, thereby embodying inherent and diverse tunabilities in both band topology and superconductivity. Investigation of the mutual couplings between these three intertwined concepts offers rich opportunities for exploring emergent physics and functionalities.

Topology has recently been introduced into condensed matter physics to describe the electronic, phononic, and superconducting properties of crystalline solids, substantially enriching our understanding of materials. For electronic states, if the wave function of a system cannot be adiabatically connected to its trivial atomic limit, the system is classified as topologically nontrivial, and is characterized by the emergence of robust boundary states protected by topological phases in the bulk [1]. Here, the topological number $Z_2$ and Chern number $C$ are employed to categorize the respective topological properties in time-reversal invariant and time-reversal symmetry breaking systems. In particular, $Z_2$ represents a global property of the occupied bands in the entire Brillouin zone, and the electronic structure of a $Z_2$ topological insulator is often accompanied by a band inversion that involves the switching of bands with opposite parities around the Fermi level. An inversion-symmetric crystal is topologically nontrivial only if the band inversion between states with opposite parities takes place at an odd number of time-reversal invariant momenta [1]. In inversion symmetry breaking crystals, $Z_2$ is directly determined by the Berry curvature and Berry connection [2]. Trivial insulators with conduction band minimum and valence band maximum located at time-reversal invariant points can be turned into topological insulators by modulating the band gap and the strength of spin-orbital coupling (SOC) via pressure/strain [3,4] or alloying with heavier elements [5,6].

Ferroelectrics are materials hosting spontaneous electric polarization states that can be reversed by an external electric field. In traditional ferroelectric interfacial devices, the macroscopic polarization field can modulate carrier concentrations and further tune various physical properties such as superconductivity, magnetism, and insulator-to-metal transition [7]. In the modern theory of polarization, electric polarization is no longer viewed as a static electric dipole (in analogy to a magnetic dipole), but as a dynamic spatial distribution of current density. The electric polarization vector is proportional to the Berry phase that is defined by the integral of the Berry curvature over the entire Brillouin zone [8]. The Berry curvature vanishes at any momentum in the Brillouin zone when both the time-reversal and inversion symmetries are preserved. However, since inversion symmetry is broken in ferroelectrics, non-vanishing Berry curvature is allowed, which may bring abnormal influences on transport and thermodynamic properties of materials. Although both ferroelectricity and topology (namely, the $Z_2$ topological number) are inherently connected to the Berry curvature, their mutual inner connection was not revealed until recently [9-15].

Among the ferroelectric materials, two-dimensional (2D) ferroelectrics have surged recently, such as in-plane polarized SnTe [16], out-of-plane polarized $CuInP_2S_6$ [17], and intercorrelated in-plane and out-of-plane polarized $In_2Se_3$ [18,19]. In addition to ferroelectric materials with strong interlayer bonds, vertical polarization can be induced in 2D van der Waals (vdW) bilayers and multilayers upon inequivalent interlayer stacking, even if the individual monolayers are non-ferroelectric, a phenomenon referred to as sliding ferroelectricity [20]. The polarization in sliding ferroelectrics can be manipulated by sliding one monolayer with respect to another, driven by a vertical electric field [20-22]. Very recently, it has been uncovered that the Berry phase, rather than symmetry, plays a central role in generating nonzero polarization in sliding ferroelectrics [23]. As highlighted in several recent reviews on 2D ferroelectric materials [24,25] and sliding ferroelectricity [26], the utilization of 2D materials and vdW stacking configurations offers unprecedented tunability and emergent phenomena, especially in the field of topological states. The inert surfaces and absence of dangling bonds in 2D materials ensure that vdW stacking of 2D monolayers in real space can directly combine the energy bands of separate monolayers in momentum space. Furthermore, the out-of-plane polarization can modify charge transfer and band alignment, thereby tuning orbital hybridization and band inversion, which can lead to the switching of non-magnetic or magnetic band topology [9-12], anomalous nonlinear Hall effect [13,15], layer Hall effect [27], etc. The breaking of inversion symmetry and ferroelectric displacement also activate extra mixing of the orbitals, resulting in an abnormally large Berry curvature dipole, which could give rise to anomalous transport phenomena such as the nonlinear Hall effect [13-15].

Superconductivity is another well-known macroscopic quantum phenomenon characterized by zero resistivity and the expulsion of magnetic field below a critical temperature. In conventional superconductors, electron-phonon coupling provides an effective attraction between the conduction electrons [28,29]. The coupling strength is intimately linked with factors such as carrier density, phonon frequency, etc. Superconductivity can also develop without the involvement of phonons, which is classified as unconventional superconductivity. In this case, the pairing mechanism does not rely on the exchanging of phonons but rather on the exchange of other bosonic excitations, such as spin fluctuations [30]. In the language of renormalization, although the bare interaction between electrons is inherently repulsive, their effective interaction can become attractive within certain parameter ranges in solids via intricate renormalization processes [31,32]. To date, a universal theory for unconventional superconductors has not been established yet. Nevertheless, the crucial role of carrier density in affecting superconductivity is widely accepted in the field. Beyond understanding the pairing mechanisms, exploration of superconductivity-related new concepts and technological applications stays at the forefront of condensed matter physics and materials science. For instance, superconducting diodes are indispensable as non-dissipative circuit elements enabling emergent superconducting technologies [33], while topological superconductors serve as a promising platform for realizing topological quantum computing [34].

As for the interplay between ferroelectricity and superconductivity, in the early stages, ferroelectric tuning of superconductivity was realized by utilizing ferroelectric/superconductor heterostructures. One pioneering experimental demonstration of ferroelectricity-modulated superconductivity was reported in 1999 using heterostructures consisting of ferroelectric $Pb(Zr_xTi_{1-x})O_3$ and superconducting $GdBa_2Cu_3O_{7-x}$ thin films [35]. In 2006, a complete switching between zero-resistance superconductivity and the normal state was realized by ferroelectric polarization reversal at a well selected temperature, using a single crystal film of the perovskite superconductor Nb-doped $SrTiO_3$ as the superconducting channel and ferroelectric $Pb(Zr,Ti)O_3$ as the gate oxide [36]. Utilizing heterostructures that combine a large-polarization ferroelectric ($BiFeO_3$) and a high-temperature superconductor ($YBa_2Cu_3O_7$), ferroelectric modulation of the superconducting condensate was demonstrated at the nanoscale in 2011; moreover, due to the materials choice and optimization, significant modulation of superconducting transition temperature was achieved, reaching magnitudes as large as 30 K [37].

The emergence of 2D ferroelectrics [16-20,38], 2D superconductors [39-41], and their vdW stacking empowers more diverse and exotic tunabilities. Recently, the ferroelectric reversed superconducting diode effect was proposed using monolayer $CuNb_2Se_4$ as a model system [42], and simultaneous tuning of band topology and superconductivity was proposed using the $IrTe_2/In_2Se_3$ heterobilayer as a concrete example [43,44]. Subsequently, theoretical proposals of ferroelectric topological superconductors [45] and ferroelectric topological nodal-point superconductors [46] have also been presented.

Another fundamental issue of ferroelectricity and superconductivity is their coexistence. Ferroelectricity and metallicity (or superconductivity) are commonly viewed to be incompatible due to the screening of long-range Coulombic interaction between dipoles by high-density itinerant electrons. To resolve this dilemma, it was suggested that a ferroelectric-like phase transition could take place in a metal if the metallic electrons interact only weakly with the ferroelectric phonons [47]. This intriguing concept was proposed over 50 years ago, but it was not until 2013 that the first unambiguous polar metal $LiOsO_3$ was discovered experimentally [48]. Coexistence of metallicity and ferroelectric-like state was also reported at low carrier doping levels in $SrTiO_3$ [49]. However, due to the three-dimensional metallic character, the reversal of polarization by an electric field, i.e., ferroelectricity, has not been confirmed in either $LiOsO_3$ or $SrTiO_3$. A truly switchable ferroelectric metal did not appear until the emergence of 2D ferroelectricity.

The vdW spacing in 2D homobilayer or heterobilayer offers a natural avenue for realizing the "weak electron-lattice coupling principle"[47]. Furthermore, sliding ferroelectricity provides a universal strategy to achieving ferroelectric states in almost any non-centrosymmetric 2D layer with various physical properties, including magnetism, metallicity, and superconductivity [20]. In addition, when a metal is thinned down to atomic thickness, an electric field can sufficiently penetrate and switch its polarity. Truly switchable spontaneous out-of-plane polarization in a metal was reported very recently in 2018 in bi- and tri-layered [50] and even bulk-like $WTe_2$ [51]. Here, it is the vdW spacing layer that enables the coexistence of semi-metallic and ferroelectric properties in $WTe_2$. The weak coupling allows the unequal distribution of charge along the out-of-plane direction (for strong interlayer coupling, non-polarized states always have the lowest energy), which protects the out-of-plane polarization, whereas free carriers are only allowed to move in the in-plane direction within each layer.

It is conceivable that we can obtain ferroelectric superconductors if the 2D building blocks of the ferroelectric systems are themselves superconductors. Indeed, two research groups have independently observed the coexistence of ferroelectricity and superconductivity in such systems, including magic-angle twisted bilayer graphene with aligned boron nitride layers [52] and bilayer MoTe$_2$ [53]. Furthermore, ferroelectricity-tuned superconducting transition temperatures as predicted in several recent theoretical studies have been experimentally observed in those systems as well [52,53].

Given the appearances of several comprehensive reviews on 2D ferroelectrics focusing on materials, ferroelectric origin, and applications [22,25], in this mini-review, we choose to focus on the couplings between ferroelectricity, electronic band topology, and superconductivity, paying special attention to fascinating tuning capabilities of ferroelectricity endowed by 2D ferroelectrics and their heterostructures. We first present results from theoretical studies of the diverse couplings between ferroelectricity, band topology, and superconductivity. Next, we cover the compelling advances on experimental observations of coexistence of ferroelectricity and superconductivity, as well as ferroelectricity-tuned superconductivity achieved in different vdW materials. Finally, we provide some perspectives towards future directions in this active research area. We hope this review will shed some light on a comprehensive understanding of the interplay between the three cornerstone concepts, and will further stimulate substantial research activities.

## 2. Theoretical Predictions of Ferroelectric Tuning of Band Topology and Superconductivity
### 2.1 Ferroelectric tuning of band topology

The electrostatic potentials on opposite sides of a 2D ferroelectric insulator with finite out-of-plane polarization are different, regardless of the existence of in-plane polarization. When it is combined with a trivial 2D insulator by vertical vdW stacking to form a heterostrcuture, the band alignment between the 2D ferroelectric and trivial insulators can be changed upon polarization reversal. Depending on the work function values of the two layers and their relative difference, the band arrangement can be varied diversely. One common scenario is the type-II band alignment, where the valence band maximum and conduction band minimum stem from different materials, facilitating ferroelectric control of topological band order associated with the interlayer hybridization and SOC. For example, in one polarization state, a significant energy gap is present between the valence band maximum from the non-ferroelectric layer and the conduction band minimum from the ferroelectric layer, such that the system is a trivial semiconductor. For the opposite polarization, the band gap is narrowed to be small enough such that a band inversion takes place with the inclusion of the SOC, driving the heterostructure into a $Z_2$ nontrivial topological state, as pictured in **Figure 1**a [11]. The switching between $Z_2$ trivial and nontrivial insulators has been theoretically demonstrated in various 2D heterostructures including Bi(111)-bilayer/In$_2$Se$_3$ [9], antimonene/In$_2$Se$_3$ (**Figure 1**b) [10], and CuI/In$_2$Se$_3$ [11]. Furthermore, it was demonstrated that the nontrivial band topology in CuI/In$_2$Se$_3$ is resilient against relative orientation and lattice mismatching between the two constituent layers [22]. Notably, with an appropriate combination of materials, $Z_2$ nontrivial insulators can also be realized at both polarizations, a phenomenon that has not been reported either experimentally or theoretically yet. In addition to heterostructures or heterobilayers consisting of ferroelectric and trivial insulators, a ferroelectric homobilayer itself can also host ferroelectrically tunable band topology, as long as the conduction band minimum and valence band maximum are located exactly or near the time-reversal invariant point (e.g., the Γ point for the In$_2$Se$_3$ monolayer). Due to the interlayer polarization-dependent band alignment and band hybridization, an In$_2$Se$_3$ bilayer exhibits polarization-dependent band topology. To be more specific, the interlayer ferroelectric configuration possesses nontrivial band topology, whereas the antiferroelectric interlayer configuration is topologically trivial [54,55].

The topological band theory has also been expanded to time-reversal breaking systems in classifying the topological properties of magnetic materials [12]. For example, due to the difference in work function and band gap, the typical type-III band alignment can be formed in a MnSe/In$_2$S$_3$ heterobilayer, which gives rise to two spin-dependent quasi-Dirac cones near the Dirac point in the absence of SOC. When including SOC, a band gap of 18 meV opens at the Dirac cone. This feature implies the existence of nontrivial band topology in the MnSe/In$_2$S$_3$ heterobilayer, which has been further demonstrated to be present in the antiferromagnetic quantum spin Hall phase. By reversing the electric polarization, the MnSe/In$_2$S$_3$ heterobilayer becomes a wide band-gap insulator, indicating a nontrivial-to-trivial phase transition (**Figure 1**c) [56]. The $Z_2$ topological insulators can be called the first-order topological insulators, while higher-order topological insulators do not exhibit gapless states in the (d-1)-dimensional boundary, but possess gapless hinge or corner states in the (d-n)-dimensional boundary with n>1. [57,58]. Furthermore, ferroelectric polarization reversal is able to switch between the $Z_2$ topological insulator and higher-order topological insulator phases. By varying the energy offsets of the ferroelectric and semiconducting layers, the heterobilayer can transform from a $Z_2$ topological insulator to a higher-order topological insulator (**Figure 1**d). Remarkably, based on an analysis of Wannier charge centers, edge states, and corner states, the ferroelectric heterobilayer composed of $MgAl_2Se_4$/$In_2S_3$ serves as a candidate material for realizing a ferroelectric higher-order topological insulator [59].

Besides the tunable band topology in ferroelectric-based heterostructures or homobilayers, the modern theory of polarization based on Berry phase is rooted in an inherent and hidden connection between ferroelectricity, band topology, and quantum geometrical phenomena in a single ferroelectric monolayer [60]. Higher-order topological insulators and corner states have been demonstrated in ferroelectric monolayers with spontaneous in-plane polarization [61]. An efficient way has also been proposed for engineering the nontrivial corner states by ferroelectricity in 2D multiferroic second-order topological insulators [61], and the underlying principle is that, changing the polarization direction can simultaneously change the positions of the corner states [61]. In addition to the aforementioned global topological properties, ferroelectricity also plays a significant role in local quantum geometrical phenomena. Generally speaking, non-centrosymmetric materials often host non-vanishing Berry curvature dipoles. Recently, a ferroelectric anomalous nonlinear Hall effect was demonstrated both theoretically [15] and experimentally [13] in ferroelectric $WTe_2$. The sign of the ferroelectric anomalous Hall current or voltage can be reversed in odd-layer $WTe_2$ (beyond the monolayer limit) by an out-of-plane electric field, which is attributed to the Berry curvature dipole reversal and shift dipole reversal caused by polarization reversal. Another more subtle and intriguing phenomenon is that ferroelectricity can enable unexpectedly large Berry curvature in a trivial wide-band-gap semiconductor [14], as it was previously believed that large Berry curvature dipoles are inherited in topological materials that possess narrow inverted band gaps or band crossings [62,63]. Appearance of large Berry curvature dipoles in wide-band-gap materials is nearly impossible because a large band gap impedes singular band inversions and crossings. Nevertheless, it has been predicted that two neighboring Berry curvature peaks with opposite signs develop and gradually move away from each other with increasing polarization magnitude, because the ferroelectric displacement activates nearest-neighboring interorbital hopping channels that efficiently mix the orbital characteristics of the conduction and valence bands [14]. Consequently, charge and spin circular photogalvanic currents can be generated in a controllable way determined by the photon handedness and ferroelectric polarization [14].

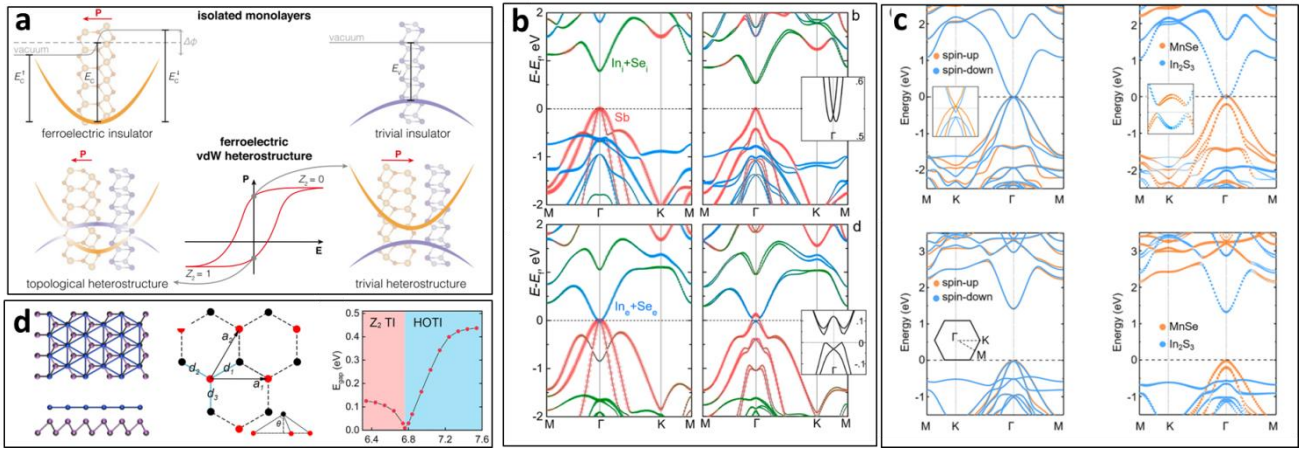

**Figure 1.** Ferroelectric switching of band topology. a) Operating principles of ferroelectric switching of band topology. Reproduced with permission. [11] Copyright 2022, Springer Nature. b) Switching between a trivial insulator and a first-order $Z_2$ topological insulator. Reproduced with permission. [10] Copyright 2021, American Chemical Society. c) Switching between an antiferromagnetic trivial insulator and an antiferromagnetic first-order $Z_2$ topological insulator. Reproduced with permission. [56] Copyright 2022, American Physical Society. d) Switching between a first-order and a second-order $Z_2$ topological insulator. Reproduced with permission. [59] Copyright 2023, American Physical Society.

**2.2 Ferroelectricity-tuned superconducting diode effect**

Superconducting diode effect, which is of scientific and technological significance, was first proposed in 1996 [64], and rises to be an active topic of research in the past two years [65-67]. The superconducting diode effect refers to the nonreciprocity of dissipationless superconducting current, i.e., the superconducting current flows in one direction, but is resistive in the opposite direction, and such behaviors can be realized in both Josephson junctions [65,68,69] and junction-free superconductors [70-72]. In principle, the superconducting diode effect may be further generalized to situations where the superconducting current exhibits imbalance between the two directions . An overview assessment on the diverse platforms and rich physics regarding recent experimental progresses and theoretical developments for the superconducting diode effect can be found in Ref.

[33]. Here we highlight the ferroelectricity-enabled and tuned superconducting diode effect in junction-free configurations.

The discovery of 2D sliding ferroelectricity [20] enables the coexistence of ferroelectricity and metallicity/superconductivity (the corresponding mechanisms will be discussed in **Section 3**) and empowers unprecedented tunability of vector quantities which is absent in 3D ferroelectrics. In a noncentrosymmetric superconductor, the critical currents along opposite directions are different, i.e., $j_c(n) \neq j_c(-n)$, where $j_c(n)$ represents the magnitude of the critical current along direction $n$. This nonreciprocity is the inherent origin of the superconducting diode effect, where the system is superconducting in one direction but resistive in the opposite direction if the applied current has a magnitude between $j_c(n)$ and $j_c(-n)$. Based on the construction of sliding ferroelectrics and a well-known 2D superconductor NbSe$_2$, a ferroelectric superconductor with reversible superconducting diode effect was proposed in Cu intercalated bilayer NbSe$_2$, with ferroelectricity induced by interlayer sliding [42]. The VI and VIII structures shown in **Figure 2**a have the same energy, but are downward and upward polarized, respectively, which gives rise to ferroelectricity. In the VI structure with downward polarization (P$_{dw}$), the energy bands across the Fermi level carry opposite spin polarizations in the +K and -K valleys, as dictated by time-reversal symmetry. In the VIII structure with upward polarization (P$_{up}$), the spin splitting in a given valley is opposite compared to that in the P$_{dn}$ case. Therefore, the spin-valley coupling is controlled by the layer polarization, which is, in turn, controlled by the ferroelectricity (**Figure 2**b). The angle-dependent critical current j$_c$(θ) is generally different for opposite directions since j$_c$(θ) ≠ j$_c$(θ + π) for a generic angle θ, as shown in **Figure 2**c. In the P$_{up}$ state, j$_c$(0) > j$_c$(π), and the system is superconducting with a current along the +*x* direction, but resistive along the -*x* direction, which leads to the superconducting diode effect. In the P$_{dw}$ state, the superconducting direction is changed to the -*x* direction with a current whose magnitude is the same as that in the P$_{up}$ state. Therefore, the superconducting diode effect is reversed upon ferroelectric polarization reversal.

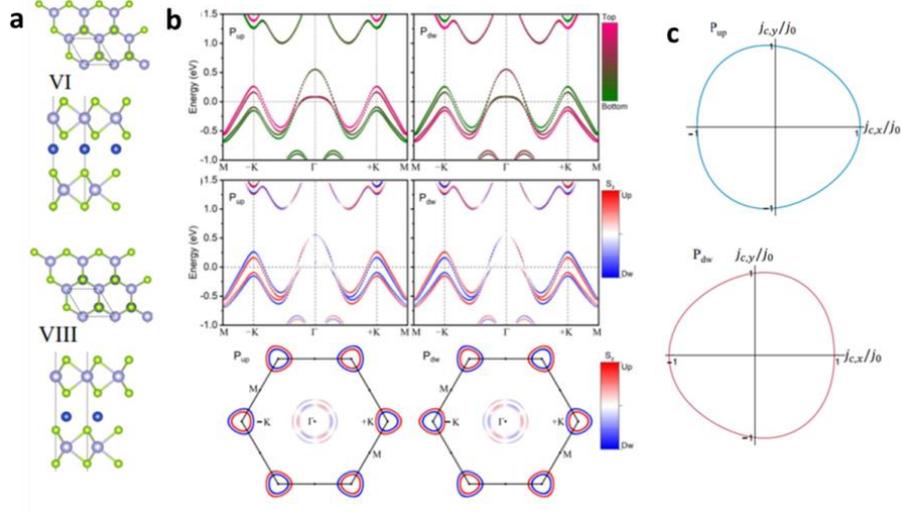

**Figure 2.** Reversible superconducting diode effect in ferroelectric superconductors. a) Top and side views of atomic structures of Cu intercalated bilayer $NbSe_2$, with VI denoting the structure with downward polarization ($P_{dw}$) and VIII denoting the structure with upward polarization ($P_{up}$). b) Layer- and spin-projected band structures of Cu intercalated bilayer $NbSe_2$. c) Angle dependence of the critical current in the $P_{up}$ and $P_{dw}$ states. Reproduced with permission. [42] Copyright 2022, American Physical Society.

## 2.3 Ferroelectric tuning of topological superconductivity

A superconductor is classified as topologically nontrivial if its wave function cannot be adiabatically transformed to a trivial Bose-Einstein condensate of Cooper pairs [73], in analogy to the definition of topological insulators. An intrinsic topological superconducting phase with odd-parity pairing is extremely rare in nature. Alternatively, researchers have resorted to an effective odd-parity pairing gap induced extrinsically by the interplay of magnetic field, *s*-wave pairing gap, and spin-momentum-locked states. The effective pairing can be realized by real-space proximity effect and reciprocal-space proximity effect [74]. The former case includes *s*-wave superconductor/topological insulator heterostructures exploiting the spin-momentum-locked edge states [75] and topologically trivial semiconducting thin films with chiral spin textures (equivalent to the edge states of topological insulators) sandwiched between a conventional *s*-wave superconductor and a ferromagnet [76,77]. The latter case is referred to superconductors with nontrivial $Z_2$ topology characterized by Dirac-type surface states, with $FeTe_{1-x}Se_x$ [5,78] as a well-known example. It is also worth mentioning that a nontrivial topological band structure is not a prerequisite for topological superconductivity, and the presence of $Z_2$ nontrivial band topology in a superconductor does not ensure a topological superconductor either. Going beyond the single particle picture, topological superconductivity can also arise from many-electron correlation effects. The delicate interplay between superconductivity, band topology, and topological superconductivity can be found in a recent review [74]. In previous studies, the manipulation of Majorana modes relies on magnetic field or gating [74]. Recently, ferroelectricity, which is superior to other tuning methods, has been introduced to tune topological superconductivity or Majorana modes [43,45,46].

More specifically, it was demonstrated that the superconducting transition temperature and band topology of a 2D $IrTe_2$ superconductor can be simultaneously tuned via proximity coupling with a ferroelectric $In_2Se_3$ monolayer [43]. The $In_2Se_3$ and $IrTe_2$ monolayers are chosen because the conduction band minimum in $In_2Se_3$ and the "valence band"(in the sense of curved chemical potential) maximum in $IrTe_2$ are both located at the center of the Brillouin zone, fulfilling the prerequisite for topological phase transition. The interlayer distance in the heterostructure shortens from 2.67 to 2.54 Å when the polarization changes from upward to downward, accompanied with a larger binding energy and increased charge transfer in the downward case. In previous studies, an $IrTe_2$ monolayer grown on graphene exhibits large atomic reconstructions [79]. However, it is found that the $IrTe_2$ monolayer is dynamically and thermodynamically stable on the ferroelectric $In_2Se_3$ layer. Figure **3**a shows the projected band structures of the $IrTe_2/In_2Se_3$ heterobilayers for both polarizations with and without the SOC. Specifically, without the SOC, there exists a small gap of ~55 meV between the $In_o+Se_o$ and Te-$p$ bands around the Γ point for the downward polarization. When the SOC is included, the $In_o+Se_o$ and Te-$p$ bands are inverted by crossing the "curved chemical potential", which may be accompanied by a topological phase transition. In contrast, for the upward polarization, there is a global gap above the Fermi level, approximately 0.93 eV around the Γ point. The SOC can reduce the gap, but cannot close it to induce a band inversion. Calculated $Z_2$ numbers confirm the nontrivial and trivial characters of the heterostructures with downward and upward polarizations, respectively. According to the bulk-boundary correspondence, a pair of topological edge states with Dirac nature at the $\bar{X}$ point is also observed within the bulk gap for the downward case, while absent for the upward case (**Figure 3**a). By invoking the proximity effect between the topological edge states and 2D bulk superconducting states [78], it is plausible that the edge of the heterostructure hosts quasi-1D topological superconducting states, which can be further exploited to harbor Majorana zero modes [75,80]. For example, Majorana zero modes can emerge at the junction between the downward polarized domain and a magnetic insulator [80]. The creation and manipulation of Majorana zero modes (or end modes) that in principle can be created at the boundaries of topologically nontrivial and trivial domains are shown in **Figure 3**b. Clear detection of Majorana modes requires avoided interference from trivial metallic states at the Fermi level [81]. In this regard, alloying $IrTe_2$ with Pd in $IrTe_2/In_2Se_3$ serves as a feasible solution, since the trivial metallic states gradually move away from the Fermi level and the Dirac point in the edge states approaches the Fermi level, as directly shown in **Figure 3**c and discussed in Ref. [44].

Another important issue is whether the ferroelectricity survives when $In_2Se_3$ binds together with the metallic $IrTe_2$ monolayer. First, the ferroelectricity in $In_2Se_3$ is dictated by the intralayer covalent bonding [18,19], which provides resistance against external perturbations induced by the vdW interfacial coupling, as evidenced by the observation of a ferroelectric hysteresis loop in $In_2Se_3$ grown on conductive substrates [19,82]. Second, first-principles calculations show that the flipping barrier is highly asymmetric, with 0.114 (0.095) eV for reversing downward to upward polarizations and 0.019 (0.015) eV for reversing upward to downward polarizations, using the PBE-D2 (vdW-optB86b) schemes (see **Figure 3**d), which is slightly higher than that of freestanding monolayer $In_2Se_3$ (0.07 eV) but in a practically reversible range.

In traditional ferroelectric/superconductor heterostructures, tuning of superconductivity is ascribed to the modulation of carrier density [4-6]. In contrast, in the 2D ferroelectric/superconductor heterobilayer of $IrTe_2/In_2Se_3$, the interlayer couplings serve as an absolute predominance in tuning the superconductivity. Based on the branch-resolved Eliashberg function $α^2F(ω)$ (shaded regions) and cumulative frequency-dependent coupling $λ(ω)$ in **Figure 3**e, it is revealed that the difference of electron-phonon coupling upon polarization reversal is mainly associated with interlayer vibration modes. It is worth emphasizing that the role of vdW interlayer coupling is pivotal, on two aspects. On one hand, the vdW interaction is strong enough for ferroelectric polarization to influence the superconducting layer (tuning of electron-phonon coupling, band alignment, and orbital interactions). On the other hand, the vdW interaction is weak enough to protect the ferroelectric layer from being destroyed by the metallic overlayer [44].

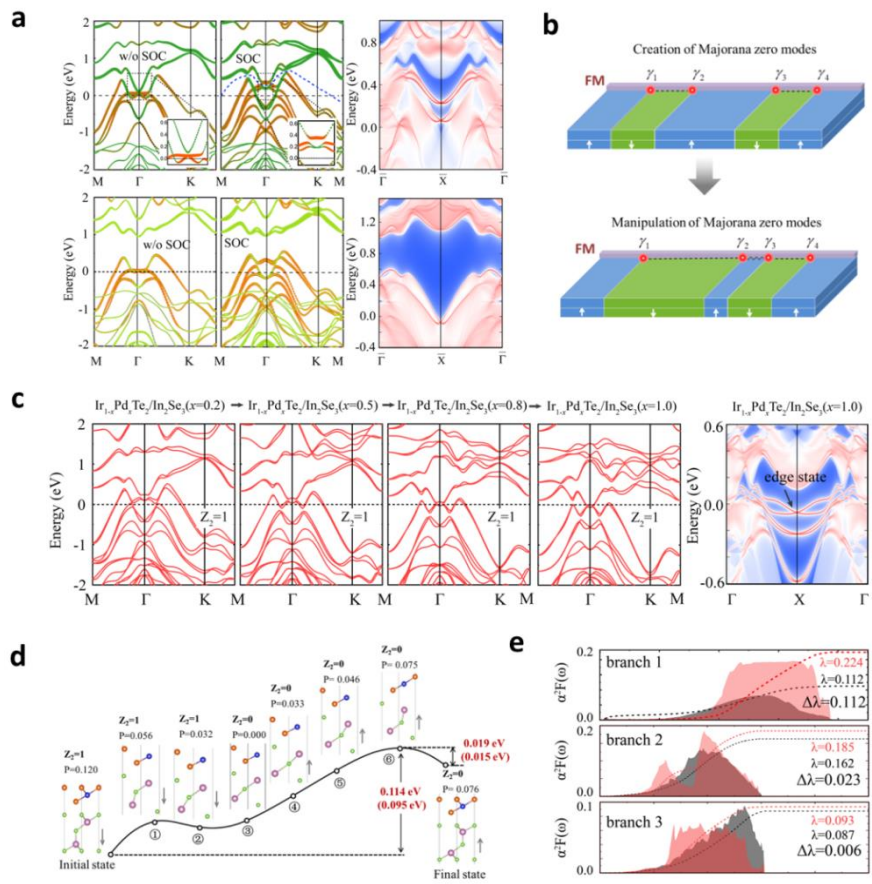

**Figure 3.** Ferroelectricity-tuned topological superconductivity based on topological edge states. a) Band structures and topological edge states in the IrTe$_2$/In$_2$Se$_3$ heterobilayers. b) Schematic diagrams of creating and manipulating Majorana zero modes in IrTe$_2$/In$_2$Se$_3$ heterostructure-based platforms. Representative topological superconductor platforms with tunable dispersive Majorana modes. c) Band structures for the downward polarized Ir$_{1-x}$Pd$_x$Te$_2$/In$_2$Se$_3$ calculated with the SOC and edge states of a semi-infinite slab of Ir$_{1-x}$Pd$_x$Te$_2$/In$_2$Se$_3$ with $x = 1.0$. d) Kinetic pathway along the trajectory defined by the electric polarization reversal for IrTe$_2$/In$_2$Se$_3$. e) Branch-resolved $\alpha^2F(\omega)$ (shaded regions) and cumulative frequency-dependent coupling $\lambda(\omega)$ (dashed lines) for the downward (red) and upward (dark gray) polarizations. Reproduced with permission. [43,44] Copyright 2023, American Physical Society.

Besides topological edge states, spin-momentum-locked states can also be utilized to generate topological superconductivity. Domain-wall dispersive Majorana modes can arise at the boundary between two topological superconductor domains with opposite Chern numbers or with a π-phase shift in their pairing gap (**Figure 4**a) [83]. A topological superconducting platform realized by a ferroelectric layer with the proximity effects of a ferromagnetic layer and superconducting substrate has been proposed [45]. Here, the chirality of the SOC plays the same role as the π-phase shift of *s*-wave pairing in changing the sign of odd-parity pairing, because the spin-momentum-locked states have also a sense of sign to distinguish their chiral spin textures. For a given material or an existing topological superconducting platform, the spin texture is generally fixed. In contrast, ferroelectric topological superconductivity enables the sign of odd-parity pairing to be easily switched by reversing the chirality of the Rashba SOC via ferroelectric polarization. Together with the sign of Chern number *C* controlled by ferromagnetic polarization, there can be four different topological superconductivity domains with different sign combinations of odd-parity pairing and *C* (**Figure 4**a) [45]. Using first-principles calculations, $In_2Se_3$ in proximity with a ferromagnetic layer and a superconductor substrate can be converted into a ferroelectric topological superconductor (**Figure 4**b).

The tunable topological superconductivity based on superconductor/ferroelectric heterobilayers relies on the switching of $Z_2$ band topology, i.e., the exploited spin-momentum locking is tied to the topological edge states [43,44]. The ferroelectric topological superconductor is rooted in the ferroelectric polarization-induced switchable chirality of spin textures arising from the Rashba SOC, which is irrelevant with band topology [45]. In fact, the noncentrosymmetric crystal structure of a ferroelectric material can induce other forms of antisymmetric SOC. By considering a ferroelectric with anisotropic band dispersion, ferroelectric topological nodal-point superconductivity can also be realized (**Figure 4**c) [46]. As a concrete example, the point group symmetry of ferroelectric 2D $BA_2PbCl_4$ (BA = $C_6H_5CH_2NH_3^+$) enables an anisotropic SOC effect, leading to spin splitting or degenerate electronic bands along different crystallographic directions with the spin-momentum-locking property provided by the SOC. Utilizing the unique intrinsic electronic property, it has been found that the reversal of anisotropic SOC chirality by flipping polarization can reverse the sign of the proximity-induced effective odd-parity pairing, leading to ferroelectric topological nodal-point superconductivity (**Figure 4**d) [46]. Consequently, by modulating the polarization direction, gapless Majorana modes can emerge at the domain wall between two oppositely polarized ferroelectric domains.

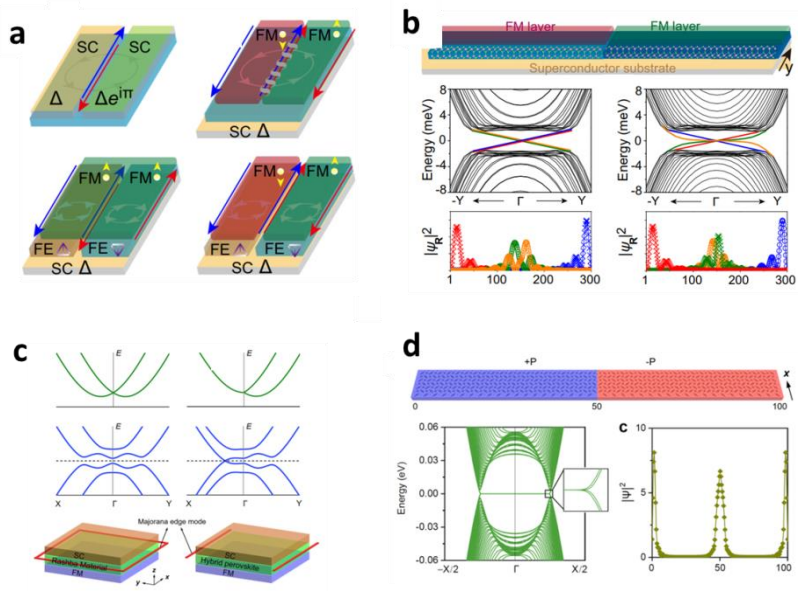

**Figure 4.** Ferroelectricity-tuned topological superconductivity based on spin-momentum-locked bands. a) Different types of Majorana modes tuned by the superconducting phases, magnetic field, and ferroelectric polarization field. b) Characterization of an $In_2Se_3$-based ferroelectric topological superconductor. Reproduced with permission. [45] Copyright 2024, American Physical Society. c) Schematics of topological superconductivity vs topological nodal-point superconductivity based on noncentrosymmetric semiconductors with isotropic vs anisotropic SOC, respectively. d) Domain-wall Majorana modes in a $BA_2PbCl_4$ monolayer. Reproduced with permission. [46] Copyright 2024, American Chemical Society.

## 3. Experimental studies of ferroelectricity-tuned superconductivity

The functionalities enabled by ferroelectric polarizations are extremely attractive, but are primarily predicted on the theoretical side. So far, direct experimental verification of ferroelectricity-tuned band topology, superconducting diode effect, and topological superconductivity has yet to be reported. But excitingly, significant experimental advancements have been achieved in confirming the coexistence of ferroelectricity and superconductivity and ferroelectric tuning of the superconducting transition temperatures in 2D materials, as elaborated in the following paragraphs [53,84].

$MoTe_2$ is isostructural to the ferroelectric metal $WTe_2$ [50,51] with both electron and hole carriers. It has been reported that the superconducting transition temperature ($T_c$) shifts up to 7 K when $MoTe_2$ is prepared as a single layer of atoms, and 2 K when it exists as bilayers [85]. The ferroelectric switching behavior in bilayer MoTe$_2$ has also been observed with an applied displacement field [53], similar to WTe$_2$. When the polarization is switched by the external electric field, the material suddenly turns from a superconductor with zero resistance into a normal metal with non-zero resistance (**Figure 5**a-c) [53]. Furthermore, MoTe$_2$ transforms from a superconductor into a normal metal upon application of an electric field pulse, and stays in this state indefinitely until the next pulse is applied. In both sweep directions, the maximum $T_c$ is therefore observed just before a switching event. This continuous tuning of $T_c$ before switching shows that the mechanism for the superconducting state is intimately tied to the internal electric field of the sample. The fact that superconductivity appears only when both the electron and hole pockets are present suggests that interband processes are behind the pairing mechanism [86]. The situation is reminiscent of iron-pnictide superconductors, which are also compensated semimetals. In that case, it has been proposed that the pairing interaction is enhanced by spin fluctuations. Whether ferroelectric fluctuation plays a role remains unsettled. The phase diagram shows that ferroelectricity and superconductivity compete with each other in bilayer MoTe$_2$. This behavior sharply contrasts to the recently observed cooperative relation between superconductivity and ferroelectric-like order in SrTiO$_3$ [49]. Further theoretical and experimental studies are required to reveal the origin of superconductivity, as well as the microscopic mechanism behind the coupling between superconductivity and ferroelectricity. Finally, as the two states with opposite polarizations are inversion partners in ferroelectric bilayer MoTe$_2$, persistent external field is required to break this symmetry and realize the switching between superconducting and normal states, otherwise the transition temperatures would be the same for both polarizations. Further efforts are needed to identify heterobilayer sliding ferroelectric superconductors and expand their tunability.

As mentioned in the preceding section, non-polar 2D monolayers can also be driven into ferroelectrics with out-of-plane polarization by inequivalent interlayer stacking [20]. Nevertheless, this mechanism is not applicable to mono-element systems such as a graphene bilayer, which preserves inversion symmetry at any stacking configuration. Surprisingly, switchable ferroelectricity in magic-angle twisted bilayer graphene with aligned boron nitride layers (MATBG/BN) has been observed recently, and the gate specific ferroelectricity has been ascribed to correlated electron induced interlayer charge transfer accommodated by a half-filled moiré band, rather than purely ionic displacements in conventional ferroelectrics [38]. Furthermore, in MATBG/BN, an enhanced ferroelectric polarization has also been observed even when the Fermi level is deeply embedded within the dispersive bands [87], suggesting that the ferroelectricity is not necessarily tied to the correlated electrons. To reconcile those different observations, one may take into consideration both the electron-driven interlayer charge transfer and sliding ferroelectricity, to potentially develop new theoretical understandings [87].

The tunability becomes more versatile in magic-angle twisted bilayer graphene, and is mainly ascribed to the correlated electrons. For example, ferroelectric hysteresis in MATBG/BN has been observed, and reproducible bistable switching between the superconducting and metallic states has been demonstrated using gate voltage or electric displacement field (**Figure 5**d-f) [84]. In addition, correlated insulating and superconducting phases can also be switched by applying external pulses. Note that the tunable correlated insulating state is absent in $MoTe_2$. The mechanism for ferroelectricity in MATBG/BN is also different from that of $MoTe_2$. In MATBG/BN, the ferroelectric behavior is an extrinsic effect that requires aligned BN substrates, whereas MATBG by itself is not ferroelectric [38,87]. Superconductivity in MATBG originates from strong correlated electrons embedded in moiré flat bands, which are absent in bilayer $MoTe_2$. The tuning of superconductivity by ferroelectric switching is via changing the electron doping levels. In contrast, tunability of superconductivity in bilayer $MoTe_2$ relies on the combined effect of internal and external electric field, leaving the carrier concentration unchanged [53].

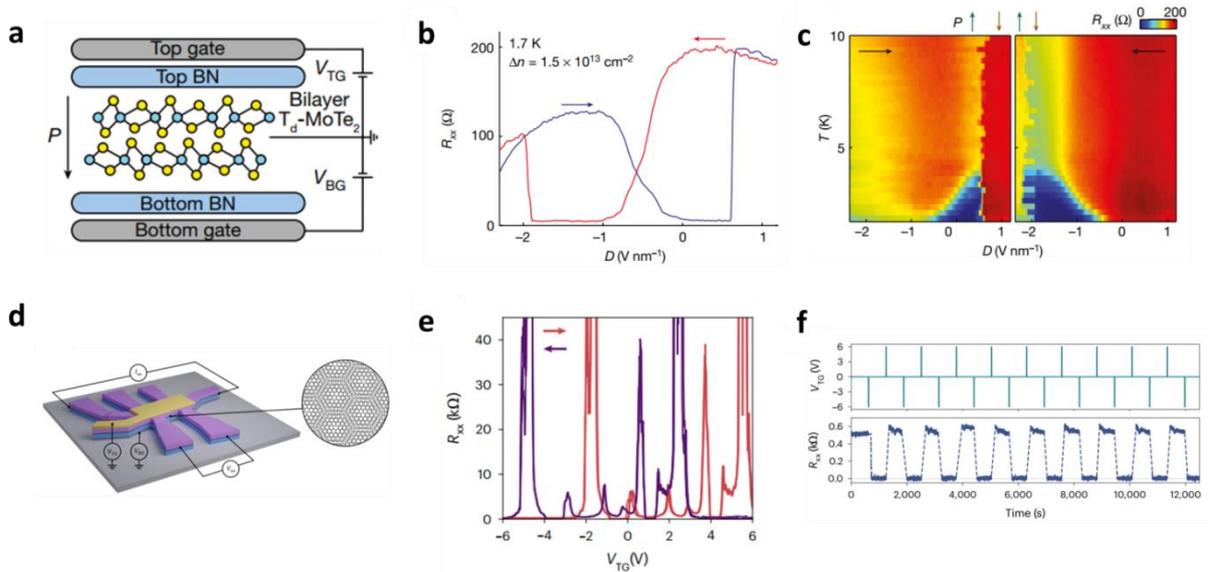

**Figure 5.** Experimentally observed reversible switching between superconducting and normal states. a) A dual-gated device based on bilayer $MoTe_2$. b) Butterfly loops with bistable normal and superconducting states, indicating coupled ferroelectric and superconducting states. The red and blue arrows denote the sweeping directions. c) Reversible switching between superconducting and normal states at a fixed carrier density, with the polarization directions denoted by the arrows. Reproduced with permission. [53] Copyright 2023, Springer Nature. d) Schematic of the dual-gated MATBG device. The inset depicts the moiré pattern of the twisted bilayer graphene heterostructure. e) $R_{xx}$ versus $V_{TG}$ with the sweeping directions up (pink) and down (purple) for $V_{BG} = 0$. f) Illustration of reversible switching of $R_{xx}$ at fixed $V_{BG} = -1.2$ V between normal metallic and superconducting states over many cycles. Reproduced with permission. [84] Copyright 2023, Springer Nature.

## 4. Discussions and Perspectives

Significant progresses have been achieved during the past few years in investigating coupled ferroelectricity, band topology, and superconductivity. The research area is still growing, and is expected to pose far-reaching influence in condensed matter physics and novel quantum devices. In this section, we give some perspective views on future likely research directions.

The advent of sliding ferroelectricity enables the coexistence of ferroelectricity and metallicity or superconductivity. The reported coupling of ferroelectricity and superconductivity has so far been limited by vdW couplings between the 2D monolayers (bilayer $MoTe_2$, MATBG/BN, $IrTe_2/In_2Se_3$ heterobilayer, Cu intercalated bilayer $NbSe_2$, etc.), and such couplings are relatively weak and external. Is it possible to find intrinsic coexistence of ferroelectricity and superconductivity in a single phase of an elemental material rather than a heterostructural system? The answer may prove to be definitive, considering that ferroelectricity can be driven by various forces, including ionic displacement, long-range magnetic order, charge redistribution, etc. [24] If so, what is the necessary ingredient for the chemical bonding within the material? What would be the interplay between ferroelectricity and superconductivity in this "strong coupling" regime? Resolving these fundamental issues is conceptually important and potentially rewarding.

Previous studies have also shown that ferroelectricity is beneficial for enhancing the superconducting transition temperature or pairing strength. $Sr_{1-x}Ca_xTiO_{3-\delta}$ (0.002<$x$<0.02) is a metal and hosts a phase transition structurally indistinguishable from the ferroelectric transition occurring in the insulator without oxygen vacancies [49]. In a small region at low doping levels, superconductivity and ferroelectricity are found to coexist. The enhancement of $T_c$ is the strongest in the vicinity of the destruction of the ferroelectric-like order [49]. It has been proposed that the soft polar phonon mode plays a major role in the formation of Cooper pairs near the peak of the superconducting dome [88]. However, an important question remains unsettled. For example, a strong enhancement of $T_c$ is expected due to the enhancement of the electron-phonon coupling close to the critical point [89]. The crucial role of ferroelectric or polar phonon modes in enhancing superconductivity is also reflected in the interface superconductor system of FeSe/SrTiO$_3$ [90]. Here, it has been experimentally found that the polar phonon modes in SrTiO$_3$ can penetrate into the FeSe overlayer with a critical thickness of two unit cells, meanwhile, these polar phonon modes are coupled strongly with the electrons in FeSe [91,92]. Strong anomalies have also been observed to be associated with a broadened ferroelectric transition in a thin layer near the FeSe/SrTiO$_3$ interface of a single-unit-cell FeSe film grown on Nb-doped SrTiO$_3$, close to the temperature at which the superconducting-like energy gap opens. The coincidence of the ferroelectric transition and gap-opening temperatures indicates the central role of the ferroelectric modes [93]. To date, a quantitative explanation of the ferroelectricity-tuned superconductivity in bilayer MoTe$_2$ is still lacking. Different from Sr$_{1-x}$Ca$_x$TiO$_{3-\delta}$ or Dirac semimetals which are near the ferroelectric quantum critical point, the ferroelectric state is quite stable in bilayer MoTe$_2$ judging from its flipping barrier. Therefore, the internal electric field may play a dominant role. More generally, we envision that by utilizing charge transfer and polar soft modes inherent in the ferroelectric/superconductor/ferroelectric interfaces, it is possible to substantially enhance the $T_c$, which may require delicate materials design and possibly high throughput screenings.

Finally, ferroelectric tuning is superior to other tuning methods such as pressure, gating, strain, or chemical doping. The ferroelectric tuning is reversible and more suitable for device applications. Different from continuous gating, ferroelectric switching does not require persistent external field to keep the bistable states, making the approach energy saving and more robust. Ferroelectric polarization-dependent topological superconductivity provides a superior knob (as compared with gating or magnetic field) for manipulation and braiding of Majorana modes, which is a crucial step towards topological quantum computing. Along this direction, further efforts are needed to search for perfect materials platforms that can avoid the subband problem and sustain large magnetic field for easy detection and practical applications. Ferroelectric tuning of other exotic quantum phenomena is also to be revealed and further enriched. More importantly, the ferroelectric tuning of electronic band topology, topological transport phenomena, superconducting diode effect, topological superconductivity, and Majorana modes represents rather appealing conceptual developments, but most of them are so far demonstrated only in theory. Future experimental studies are critically needed to validate many of these strong predictions and expectations, thereby rapidly advancing the field.

**Acknowledgements**

This work was supported by the Innovation Program for Quantum Science and Technology (Grant No. 2021ZD0302800), the National Natural Science Foundation of China (Grant Nos. 12374458, 11974323, and 12204121), the Anhui Initiative in Quantum Information Technologies (Grant No. AHY170000), and the Strategic Priority Research Program of Chinese Academy of Sciences (Grant No. XDB0510200). J.C. also acknowledges the support from the Guangxi Natural Science Foundation (Grant Nos. 2019GXNSFBA245077 and 2021GXNSFAA220129).

**Conflict of Interest**

The authors declare no conflict of interest.

**Author Contributions**

All authors made important contributions to the review.

**Keywords**

ferroelectricity; superconductivity; band topology; two-dimensional materials; van der Waals heterostructures

Received: May XX, 2024
Revised: June XX, 2024
Published online: July XX, 2024

ToC figure

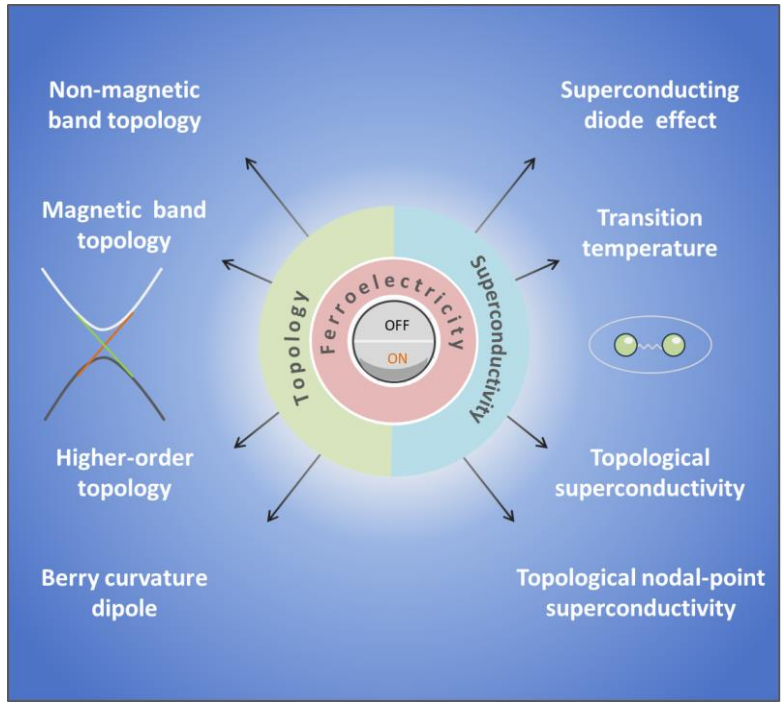